# A (meta)metafísica da ciência: o caso da mecânica quântica não-relativista[1]

Raoni Wohnrath Arroyo[2]

Jonas R. Becker Arenhart[3]

**Resumo**

Tradicionalmente, ser realista sobre algo significa crer na existência independente desse algo. Em termos ontológicos, isto é, *acerca do que há*, o realismo científico pode ser entendido como envolvendo a adoção de uma ontologia que seja cientificamente informada. Mas segundo alguns filósofos, a atitude realista deve ir além da ontologia. A forma com que essa exigência tem sido entendida envolve fornecer uma metafísica para as entidades postuladas pela ciência. Discutimos como duas abordagens em voga encaram o desafio de fornecer uma metafísica para a ciência: uma forma de naturalismo e a abordagem Viking/*Toolbox*. Por fim, apresentamos uma terceira via, que adota o melhor das duas abordagens: o método meta-popperiano, que foca em descartarmos quais as alternativas erradas, ou melhor dizendo, os perfis metafísicos incompatíveis com certas teorias. Apresentamos o método meta-popperiano, um método de metametafísica capaz de avaliar objetivamente quais os perfis metafísicos que são incompatíveis com certas teorias científicas. Para isso, usaremos como estudo de caso a mecânica quântica, mostrando resultados obtidos previamente. Com esse método, podemos ver como a ciência pode ser usada para evitar o erro em questões metafísicas. Essa seria, na nossa opinião, uma forma de desenvolver uma relação produtiva entre ciência e metafísica.

---

[1] A ordem de autoria é alfabética, e não representa qualquer tipo de prioridade. Os autores trabalharam igualmente no presente manuscrito.

[2] Centro de Lógica, Epistemologia, e História da Ciência, Universidade Estadual de Campinas. Apoio: nº 2021/11381-1, Fundação de Amparo à Pesquisa do Estado de São Paulo (FAPESP). Universidade Federal de Santa Catarina, Programa de Pós-Graduação em Filosofia, Florianópolis, SC, Brasil. Grupo de Lógica e Fundamentos da Ciência (CNPq). ORCID 0000-0002-3800-8505. E-mail: rwarroyo@unicamp.br

[3] Universidade Federal de Santa Catarina, Departamento de Filosofia, Florianópolis, SC, Brasil. Universidade Federal do Maranhão, Programa de Pós-Graduação em Filosofia, São Luís, MA, Brasil. Grupo de Lógica e Fundamentos da Ciência (CNPq). ORCID 0000-0001-8570-7336. E-mail: jonas.becker2@gmail.com

**Palavras-chave:** Metametafísica. Metafísica da ciência. Metafísica e ontologia. Mecânica quântica.

**Abstract**

Traditionally, to be a realist about something means believing in the independent existence of that something. In this line of thought, a *scientific* realist is someone who believes in the objective existence of the entities postulated by our best scientific theories. In metaphysical terms, what does that mean? In ontological terms, i.e., in terms *about what exists*, scientific realism can be understood as involving the adoption of an ontology that is scientifically informed. But according to some philosophers, the realistic attitude must go beyond ontology. The way in which this requirement has been understood involves providing a metaphysics for the entities postulated by science, that is, answering questions *about the nature* of what ontology admits to exist. We discuss how two fashionable approaches face the challenge of providing a metaphysics for science: a form of naturalism and the Viking/Toolbox approach. Finally, we present a third way, which adopts the best of both approaches: the meta-Popperian method, which focuses on discarding the wrong alternatives, or better saying, the metaphysical profiles incompatible with certain theories. We present the meta-Popperian method, a metametaphysical method capable of objectively assessing which metaphysical profiles are incompatible with certain scientific theories. For this, we will use quantum mechanics as a case study, presenting some previously obtained results. As our focus is on methodological questions about the relationship between metaphysics and science; with this method, we can see how science can be used to avoid error in metaphysical issues. In our opinion, this would be a way to develop a productive relationship between science and metaphysics.

**Introdução**

É um lugar-comum afirmarmos que a metafísica, enquanto disciplina, está em maus lençóis. Em épocas mais recentes, desde as críticas do positivismo do séc. XIX, as credenciais epistêmicas dessa área da filosofia foram colocadas em cheque de um modo que se torna cada vez mais comum: através de comparações com o sucesso da ciência empírica. Na literatura contemporânea, tem-se alegado, por um lado, que a metafísica, enquanto disciplina, falha em levar em consideração os desenvolvimentos da ciência, sendo áreas totalmente desconectadas (Callender, 2011); com esse argumento, sugere-se que temos muito pouco a aprender com a metafísica, ou que esse é o motivo pelo qual a área deve ser descontinuada (Ladyman e Ross, 2007). Por outro lado, questiona-se o que pode ser colocado no lugar da metafísica: se a própria *ciência* (Ibid.); ou se somente a *metafísica* poderia ocupar o lugar da metafísica (French, 2018a), já que essa é imprescindível para o entendimento do empreendimento científico (Ibid.; Chakravartty, 2007). Assim, a discussão sobre a "metafísica da ciência"[4] e a própria relação entre metafísica em ciência passa a ser um tópico obrigatório na metodologia da metafísica, ou metametafísica (Tahko, 2015).

Esse dilema pode ser resumido na pergunta pela credibilidade epistêmica da metafísica, que, por um lado, argumenta-se, pode herdar certas credenciais epistêmicas da ciência (Morganti e Tahko, 2017), mas, por outro lado, deve ter certa autonomia como disciplina (French, 2014; 2018). Existiria um meio termo entre um modo de fazer metafísica que, por um lado, é tão naturalizada ao ponto de simplesmente repetir a ciência relevante, e outro modo que é tão desconectado da ciência que acaba sendo irrelevante para uma descrição que é (minimamente) cientificamente informada?

Neste artigo, temos dois objetivos: 1) oferecer um panorama do atual estado da arte nesse debate e 2) apresentar uma descrição de uma alternativa até o momento pouco explorada na literatura, que parece trazer avanços para o debate. Para tanto, conduzimos essa discussão com uma articulação do realismo científico, que pode ser caracterizada na seguinte questão: o quão metafisicamente informado o realismo científico deve ser? Isso é feito na segunda seção. Adotaremos uma terminologia recente no debate sobre o realismo científico (Magnus, 2012; French 2018a), que distingue entre versões do realismo de tipo 'raso' e 'profundo'. Feito isso, analisaremos criticamente duas posições em voga, que respondem a questão de como a ciência e a metafísica devem se relacionar, enquanto disciplinas: a abordagem da naturalização da metafísica (Ladyman e Ross, 2007) na terceira seção, e a abordagem Viking à metafísica (French, 2014; 2019) — também chamada de abordagem "*Toolbox*" (French e McKenzie, 2012, 2015; French, 2018b) — na quarta seção. Na quinta seção, apresentaremos o método meta-popperiano (Arenhart, 2012; Arenhart, Arroyo, 2021a) como uma alternativa que busca o melhor dos dois mundos: a autonomia da metafísica enquanto disciplina, e o importe de credenciais epistêmicas na ciência para verificar a compatibilidade entre teorias científicas e teorias metafísicas. Concluímos na sexta seção.

---

[4] Todos os termos e citações foram traduzidos livremente pelos autores.

**Realismo científico: raso e profundo**

Um lugar bastante comum para visualizar o debate metodológico sobre a relação entre metafísica e ciência é o realismo científico. Tradicionalmente, entende-se que ser "realista" sobre algo significa manter uma crença sobre a *existência* independente desse algo; nesse raciocínio, o realismo *científico* expressaria a crença na existência independente das entidades postuladas pelas nossas melhores teorias científicas (Chakravartty, 2017). Em termos ontológicos isso quer dizer que esse tipo de realismo admite uma ontologia que seja cientificamente informada.[5]

Para muitos autores, isso é suficiente — ou, ao menos, é só até aí que devemos ir se queremos relacionar ontologia e ciência de forma a preservar um lastro científico para a nossa ontologia. Esse tipo de realismo é caracterizado como um realismo científico 'raso'; realistas que desejam ir além da ontologia, e vestir as entidades existentes segundo a teoria científica com uma camada metafísica, são realistas científicos do tipo 'profundo' (Magnus, 2012; French 2018a). A terminologia se deve a uma analogia com risco epistêmico e humildade epistêmica: realistas que preferem ficar nas águas mais rasas assumem menos risco de errar em questões metafísicas, já que ficam somente com o que é fornecido pelas teorias científicas. Assim, essa é uma atitude *epistemicamente humilde*: já que não temos garantia alguma (ao menos, não por parte da ciência) de que a metafísica possa fornecer uma descrição verdadeira da realidade, é melhor não arriscar! No entanto, o preço de ficar 'no raso' é explicar pouco, como, por exemplo, a natureza das entidades com as quais a ciência trabalha — isso seria tarefa da *metafísica* (Thomson-Jones, 2017). Essa seria a vantagem de se aventurar em águas mais 'profundas'. No entanto, o preço de reduzirmos a humildade epistêmica é aumentarmos seu risco: afinal, quanto mais profundas as águas, mais perigosas.

---

[5] Esta seção simplifica um argumento que foi desenvolvido em maiores detalhes em Arroyo e Arenhart (2020b).

A distinção entre os níveis raso e profundo é, portanto, uma distinção envolvendo análise de riscos, avaliando os prós e os contras, guiada pela questão sobre o quanto de metafísica devemos admitir numa descrição científica da natureza. Comecemos pelo lado 'raso'.

**As águas rasas da metafísica naturalizada**

A metafísica naturalizada, conforme entendida mais recentemente, é uma proposta que começa com um manifesto contra a metafísica entendida enquanto disciplina autônoma, desvinculada da ciência. Nesse manifesto, haveria dois tipos de metafísica, uma boa e outra ruim. Essa metafísica identificada como "ruim" teria diversos nomes. Alguns deles são:

- Metafísica neo-escolástica (Ross e Spurrett, 2004);
- Metafísica analítica, de poltrona, tradicional (Ladyman e Ross, 2007);
- Metafísica de fantasia (French e McKenzie, 2012);
- Metafísica baseada em intuições (Guay e Pradeu, 2020);

- Metafísica ao ar livre (Bryant, 2020a);
- Metafísica a priori (McKenzie, 2020).

Vamos assumir, por hora, que todos esses nomes se referem à mesma coisa: à metafísica enquanto disciplina, desenvolvida nos livros que estão na seção de metafísica na biblioteca, e que se qualificam como utilizando do método 'analítico' de se fazer filosofia (o que também não é exatamente claro, mas esse não é um ponto a ser discutido aqui). Vamos chamar essa metafísica de "metafísica analítica", para não introduzirmos ainda mais um nome. Eis os adjetivos que a literatura crítica do projeto atribui à metafísica analítica:

- "Irrelevante" e "pseudocientífica" (Ladyman e Ross, 2007, p. vii);
- "Fútil" (French e McKenzie, 2015, p. 28);
- "Estéril" e "vazia" (Callender, 2011, p. 34);
- "Epistemicamente inadequada" (Bryant, 2020a, p. 1867).

Um marco nessa discussão metametafísica, foi a obra de Ladyman e Ross (2007, p. vii), visto que foram eles quem introduziram a polarização no debate de maneira mais incisiva: "[…] a metafísica analítica contemporânea é uma atividade profissional desenvolvida por algumas pessoas extremamente inteligentes e moralmente sérias, [mas] não se qualifica como parte da busca esclarecida da verdade objetiva e deve ser descontinuada".

O motivo da suspeita é que a metafísica analítica é ruim justamente porque ela não se relaciona com a ciência atual. O argumento é o seguinte: a ciência tem resultados impressionantes; a metafísica não. A ciência tem progredido em nos fornecer uma descrição objetiva da realidade; a metafísica não. Logo, somente as questões que podem ser respondidas com base na ciência são questões que têm alguma credencial epistêmica. As questões que não podem ser feitas ou respondidas com o auxílio da ciência devem ser abandonadas. E essas são justamente as questões próprias da metafísica analítica — a metafísica 'ruim'.

Uma boa metafísica — a metafísica *naturalizada* — deve ser "uma visão unificada do mundo derivada dos detalhes da pesquisa científica" (Idem, p. 65). Essa atitude é frequentemente chamada de 'naturalismo'. Ainda que o termo não tenha um único entendimento na literatura, e por vezes não seja muito bem definido (Bryant, 2020b), temos em mente que 'naturalismo' é a atitude cuja essência foi destilada por Wallace (2012, pp. 3–4) na seguinte frase: "[…] a tese de que não temos melhor guia para a metafísica do que a prática bem-sucedida da ciência". No entanto, muitos naturalistas privilegiam a ontologia. A seguinte citação exemplifica isso.

> Metafísica é ontologia. Ontologia é o estudo mais genérico do que existe. A evidência do que existe, pelo menos no mundo físico, é fornecida exclusivamente pelas pesquisas empíricas. Consequentemente, o objeto apropriado da maior parte da metafísica é a análise cuidadosa de nossas melhores teorias científicas (e especialmente das teorias físicas fundamentais) com o objetivo de determinar o que elas implicam sobre a constituição do mundo físico. (Maudlin, 2007, p. 104).

O naturalismo, nessa perspectiva, é uma forma de realismo científico no qual a ciência desempenha um papel *epistemológico* central para a metafísica da ciência *enquanto* ontologia da ciência. Recusando-se a ir além de onde a ciência vai e defendendo a crença na existência das entidades com as quais a ciência se compromete, pretende substituir a metafísica *enquanto* metafísica pela própria ciência, ou, mais especificamente, pela *ontologia* da ciência. Em termos metodológicos, aspira extrair a metafísica da ciência, mas por recusar-se a ir além da ciência, repete aspectos relevantes da descrição científica e chama isso de metafísica científica (Ladyman e Ross, 2007). Mas isso não é metafísica no sentido analítico, isso é ciência. Isto é, para os naturalistas, a descrição metafísica é preenchida com o conteúdo metafísico extraído das próprias teorias científicas. Mas tomemos a física como exemplo: se a metafísica não pode ir além da física, qual metafísica é extraída da física, senão a própria… física? Por exemplo, podemos dizer que a mecânica quântica não-relativista (doravante apenas "mecânica quântica") se compromete com entidades do tipo 'elétron', entre outras; isso é *extraído* da teoria científica, pois entidades desse tipo cumprem um papel explicativo no funcionamento da própria teoria científica — isto é, desempenham papéis importantes na descrição de experimentos de decaimento de partículas. No entanto, a mecânica quântica não nos informa nada sobre o *perfil metafísico* desse tipo de entidade como, por exemplo, o "perfil de individualidade" (Branding e Skyles, 2012) — isto é, se as entidades devem ser entendidas metafisicamente enquanto *indivíduos* ou *não-indivíduos*. Como a literatura já tem mostrado ao longo dos anos, a mecânica quântica é *compatível* com perfis metafísicos de individualidade e não-individualidade, mas não nos força nenhuma das duas opções (ver French e Krause, 2006; Arenhart, 2017). Nesse sentido, fica claro que a metafísica de individualidade, para ficarmos com o caso do exemplo, é uma *adição* à teoria, e não algo que advém dela. French (1995, p. 466) já havia identificado esse problema quando afirmou o seguinte: "[…] o tanto de metafísica que você extrai de uma teoria física é o tanto que você coloca; é preciso de certos truques filosóficos para tirar coelhos metafísicos de cartolas físicas".

    Se a metafísica só é boa na medida em que se relaciona efetivamente com a ciência, ainda não sabemos ao certo o que isso significa, em quais termos essa relação deve ser entendida. Pois uma metafísica *radicalmente naturalizada* arrisca-se tão pouco, que acaba repetindo aspectos relevantes da ciência, exaurindo a metafísica de qualquer conteúdo próprio.[6] Se identificarmos metafísica e ontologia, saímos com as mãos vazias — no que diz respeito ao conteúdo metafísico. Por isso, para podermos apreciar melhor os resultados do projeto naturalista, é importante distinguir o conteúdo das duas disciplinas, 'ontologia' e 'metafísica', como faz Hofweber no seguinte trecho:

---

[6] Uma alternativa seria uma metafísica que fosse "moderadamente naturalizada", conforme oferecida por Morganti e Tahko (2017). Essa alternativa metodológica promete trazer o melhor dos dois mundos: uma descrição metafísica da realidade com garantias epistêmicas fornecidas pela ciência; no entanto, os autores falham em identificar como seria de fato uma relação produtiva entre ciência e metafísica, deixando a proposta mais como uma carta de intenções, do que uma proposta de fato. Esse assunto será melhor discutido na quarta seção, com a promessa de que uma interação de fato entre ciência e metafísica será a apresentada.

> Na metafísica, queremos descobrir como a realidade é de um modo geral. Uma parte disso será descobrir as coisas ou os materiais [*stuff*] que são partes da realidade. Outra parte da metafísica será descobrir o que essas coisas, ou esses materiais [*stuff*], são, de modo geral. A ontologia, nessa abordagem bem padrão da metafísica, é a primeira parte do projeto, i.e. é a parte da metafísica que busca descobrir quais coisas formam a realidade. Outras partes da metafísica elaboram sobre a ontologia e vão além dela, mas a ontologia é central para elas. (Hofweber, 2016, p. 13).

Então talvez possamos conceder que as teorias científicas fornecem o catálogo da realidade. Dessa forma, a ontologia, que trata sobre o que há, pode ser naturalizada no sentido de que podemos *extrair* das teorias científicas o catálogo do que existe (ou parte dele) — sempre de acordo com essas teorias em questão (ver também Arenhart e Arroyo, 2021b). Assim, se, como sugere Maudlin (2007, p. 104) na passagem destacada anteriormente, tudo o que importa é a ontologia, devemos olhar para as teorias científicas para aprender com elas o que existe.

Mas note que não há "metafísica" na metafísica naturalizada. Ao menos não na metafísica entendida enquanto o estudo da natureza das coisas, que irá nos dizer, por exemplo, se podemos individuar as entidades de acordo com um feixe de propriedades, ou se a natureza dessa individuação é um substrato que transcende suas propriedades (French e Krause, 2006; Benovsky, 2016). Se a metafísica pudesse ser descoberta na própria teoria científica, não haveria necessidade de uma investigação filosófica: a resposta para a pergunta metafísica seria apenas mais um fato científico. Não faria sentido procurar conceitos metafísicos para revestir as entidades postuladas. Esse parece ser o dilema epistêmico para a metafísica naturalizada: a metafísica deve vir de outro lugar que não a própria ciência; caso contrário, seria impossível atribuir um perfil metafísico a uma dada teoria científica. Todavia, é exatamente a relevância desse conteúdo vindo de outra fonte que é negado pelos naturalistas, tomados no sentido que estamos discutindo. O fato de não provir da ciência é o que o torna epistemicamente desqualificado.

Afinal: *quem precisa de metafísica?*

**Águas profundas para realistas científicos com aspirações metafísicas**

Como vimos, realistas científicos do tipo 'profundo' não se contentam com a imagem científica do mundo. French (2014; 2018a), por exemplo, alega que o realista científico do tipo 'raso' teria mais motivos para considerar-se um *antirrealista* do que realista científico (voltaremos a isso mais tarde), mas o que está em jogo é a alegação de que a imagem resultante da investigação exclusivamente científica pareça muito crua para justificar o próprio título de 'realismo'. Talvez uma das passagens mais representativas nesse sentido seja a seguinte, por Chakravartty (2007, p. 26): "[n]ão se pode apreciar completamente o que significa ser realista até que se tenha uma imagem clara acerca daquilo sobre o que estamos sendo convidados a ser realistas".

Segundo realistas 'profundos', devemos fornecer uma "imagem clara" acerca daquilo sobre o que dizemos ser realistas, e fornecer essa imagem clara é fornecer uma descrição metafísica das entidades acerca das quais estamos alegando ser realistas. Em outras palavras, o realismo profundo envolve responder questões acerca da *natureza* daquilo que as teorias admitem que *existe*. Realistas que não oferecem uma descrição metafísica não são realistas em um sentido legítimo: "[a]s pessoas que rejeitam essa necessidade são ou empiristas no armário ou realistas '*ersatz*'" (French, 2014, p. 50). Isso é o caso pois uma atitude empirista é definida, de modo geral, como a recusa de contrair compromisso ontológico com entidades não-observáveis, justamente por se acreditar que não podemos ter acesso epistêmico adequado a elas. Isto é, segundo essa linha de argumentação, a falta de compromisso com as entidades não-observáveis se explica pelo fato de que não sabemos o que são. Dessa forma, realistas que se recusam a ir 'fundo' fazem algo similar com as entidades não-observáveis da ciência, ao se recusarem a colocar uma camada metafísica que garante a inteligibilidade exigida. Fica faltando muito na descrição da entidade, alega-se, e isso conduz o realista raso ao mesmo tipo de situação que o empirista, só que ele não reconhece isso (porque está no armário).

Balancear a obtenção dessa imagem clara com uma quantidade certa de humildade epistêmica é o que French (2014) chama de "Desafio de Chakravartty". Isto é, se de um lado precisamos designar uma metafísica para as teorias científicas para que se possa adotar uma atitude *legitimamente* realista em relação a essas teorias, por outro lado, nós não temos um "olhar divino" para saber *se há* uma descrição metafísica adequada para as entidades postuladas por uma certa teoria científica, e *qual seria* essa descrição. Então, como fazer isso sem perder a conexão com a segurança epistêmica que a ciência supostamente nos fornece? Uma resposta é dada pela Abordagem Viking à metafísica (French, 2014; 2019), ou abordagem *Toolbox* (French e McKenzie, 2012; 2015; French, 2018b).

Basicamente, essa proposta metametafísica consiste em deixar que a metafísica analítica produza conteúdo metafísico livremente, já que esse conteúdo *pode vir a ser usado* pela filosofia da ciência (mais especificamente, pela *metafísica da ciência*) para preencher eventuais lacunas metafísicas, de modo a legitimar a adoção de uma atitude realista em relação à ciência. Em uma forma de slogan, essa abordagem diz: "sejam livres, metafísicos!" (French, 2015). Isto é, a busca por uma articulação legítima de uma forma de realismo, originadas através do Desafio de Chakravartty, sugere que se faça uso de ferramentas metafísicas que *não são* cientificamente informadas em sua origem.

A abordagem Viking/*Toolbox* não *produz* teorias metafísicas, mas usa (saqueia) as existentes — daí seu nome. Dessa forma, responde ao Desafio de Chakravartty usando teorias disponíveis na história da filosofia, ou seja, opera aplicando conceitos metafísicos existentes nas teorias científicas. A abordagem Viking/*Toolbox* fornece essa "imagem clara", indo além do que a ciência afirma. A descrição metafísica é, então, uma camada teórica adicional — de modo algum *extraída* da ciência (Arenhart, 2012; 2019; Arenhart e Arroyo, 2021b; Arroyo e Arenhart, 2019; 2020a). Assim, o proponente dessa abordagem defende um desenvolvimento irrestrito de teorias metafísicas, justificando essa liberdade criativa na possibilidade do uso das teorias metafísicas pela filosofia da ciência no futuro. De um ponto de vista metodológico, o que essa abordagem metametafísica nos diz é que ciência e metafísica operariam em níveis metodológicos completamente independentes; quando conveniente, a metafísica analítica pode ser usada para fins interpretativos na ciência, e isso seria feito por filósofos da ciência. Usando uma analogia frequente nesses contextos, os conceitos e teorias da metafísica analítica teriam um papel similar ao papel da matemática pura na formulação de teorias empíricas. Assim como teorias matemáticas que hoje não possuem aplicação empírica poderão ser usadas na formulação de teorias empíricas no futuro, teorias metafísicas sem nenhuma conexão com a ciência empírica poderão ser usadas por filósofos da ciência para articular versões profundas de realismo e dar uma imagem clara da ciência no futuro.

Nesse ponto, alguém poderia perguntar: "Mas afinal, como a Abordagem Viking/*Toolbox* responde ao Desafio de Chakravartty?", ao que podemos responder o seguinte. O metafísico da ciência escolhe, dentre as opções disponíveis na literatura metafísica (analítica), aqueles conceitos e teorias que considera apropriados para atribuir às entidades postuladas um perfil metafísico. Como a metafísica, enquanto disciplina, é independente da ciência empírica, fornece apenas as teorias, ferramentas e estratégias de investigação e especulação que podem ser usadas para interpretar teorias científicas. Cabe ao filósofo da ciência intermediar o encontro das duas, e garantir que houve, realmente, um aumento de inteligibilidade da ciência através do uso da metafísica — *e.g.*, poder saber, com a clareza necessária, sobre o que alguém está sendo realista, como sugere o Desafio de Chakravartty.

Dito de outro modo, podemos apreciar o tipo de proposta teórica feita pela abordagem Viking através da diferença entre versões 'rasas' e 'profundas' do realismo, uma diferença captada, por sua vez, através da distinção entre 'ontologia' e 'metafísica'. Em termos gerais, o realista científico do tipo 'raso' ficaria restrito ao nível ontológico, repetindo os pronunciamentos da ciência acerca de suas postulações, pela garantia epistêmica que essa restrição herda as mesmas garantias da teoria científica sendo recitada. A abordagem Viking, por sua vez, é uma forma de realismo científico profundo ou *metafísico*, justamente por ir além do nível ontológico. Vai além do que a ciência afirma, para dizer *como* é o mundo, levando em conta a ciência, mas não se reduzindo a ela e, ao fazer isso, é uma posição com menor humildade epistêmica e consequentemente com maior risco epistêmico (ela não se restringe a recitar a teoria científica relevante, como o realismo raso faz). Para responder ao Desafio de Chakravartty, utiliza teorias metafísicas disponíveis na literatura para interpretar as teorias científicas, e chama isso de metafísica da ciência.

Mas espere: como isso pode ser visto como uma resposta para a conexão produtiva entre ciência e metafísica? Isso é apenas metafísica analítica, com um nome diferente! Note: ao produzir metafísica, um metafísico analítico não precisa se restringir aos resultados da ciência, o plano é que se esteja avançando uma descrição geral do funcionamento da realidade. Ao aplicar isso em uma teoria científica, o Viking estaria apenas chamando nossa atenção para um fato já conhecido: aquela teoria metafísica está realmente dando conta de aspectos específicos da realidade. Ora, o metafísico analítico não vai se impressionar. Esse era o objetivo desde o começo!

Vamos recapitular o que foi dito para uma breve avaliação das duas propostas que consideramos até aqui. A descrição metafísica é uma descrição adicional. Caso o naturalista reconhecesse isso, ele pararia por aqui, já que recusa-se a ir além da descrição científica (nesse sentido, a metafísica científica não é metafísica, é apenas mais descrição científica). Mas o Viking iria além em sua cruzada interpretativa, "vestindo" as entidades científicas com alguma roupagem metafísica existente. No entanto, ao apenas transpor metafísicas "prontas" para o campo da ciência, falha em oferecer uma justificação para a metafísica que seja convincente para os críticos mais radicais. Para exemplificar, voltemos à questão do perfil metafísico de (não-)individualidade na mecânica quântica. O adepto da abordagem Viking estaria sugerindo, nesse caso, que podemos empregar alguma noção de individualidade, ou não-individualidade (se houver algum tal perfil), disponível na literatura metafísica e que foi desenvolvida sem se ter a mecânica quântica em mente. Mas qual delas empregar? O problema com a sugestão é que diferentes abordagens estão disponíveis, podendo ser consistentes com a mecânica quântica. Assim, a simples conexão com a teoria não nos dá nenhum motivo especial para pensar que a mecânica quântica está desempenhando um papel de atribuir algum tipo de justificação privilegiado para nossa escolha particular. No melhor dos casos, a disputa entre diferentes opções de perfis de individualidade e não individualidade se dará novamente no âmbito metafísico, não no âmbito da mecânica quântica, e *isso* era exatamente o que os críticos apontavam como insatisfatório. A metafísica da ciência, entendida através dessas lentes, não difere em conteúdo da metafísica que *não* é da ciência — só está exibida em um local diferente, buscando adquirir uma relevância que ela não tem.

Como é bem sabido, tanto o realismo científico 'raso' quanto o 'profundo' precisam conviver com o obstáculo da subdeterminação. Isso fica claro no caso da mecânica quântica: existem diversas maneiras diferentes de explicar o que ocorre com os fenômenos quânticos, e a própria física não nos fornece elementos suficientes para decidirmos por uma em detrimento das outras. Não abordaremos a subdeterminação teórica, que é uma questão a ser resolvida pela ciência (se for resolvida, de qualquer modo). No entanto, de um ponto de vista metodológico,[7] *ainda que* essa subdeterminação seja resolvida, temos outra: a metafísica. Caso admitamos uma descrição metafísica como uma camada extra, livre da física para especulação e desenvolvimento, teremos inevitavelmente uma subdeterminação da metafísica à vista: um filósofo da ciência com inclinações metafísicas *sempre* pode interpretar uma entidade (dar a ela um *perfil metafísico*) de modos diversos (e nossa exemplificação anterior com o caso da individualidade e não-individualidade ilustra exatamente isso). Já que a física não oferece ancoragem (e esse era o problema desde o começo!), a metafísica *flutua livre* da física (Arroyo; Arenhart, 2020a). Assim, não temos como saber o que de fato são as coisas com as quais a ciência se compromete.

A propaganda inicial prometia que existe uma boa metafísica, e que essa metafísica seria "boa" porque estaria relacionada à ciência. Em nossa discussão até aqui, não discutimos se isso faz mesmo algum sentido. Ao contrário, tomamos essa afirmativa literalmente, pra ver até onde ela vai (e, ao que parece, não é muito longe). Do lado do naturalismo, a "boa metafísica" não é uma metafísica ainda: ela repete a ciência, e não chega a abordar questões metafísicas, e.g. sobre a natureza das entidades da ontologia. Do lado Viking, a "boa metafísica" é exatamente a "metafísica ruim", mas com outro nome: *metafísica da ciência!*

---

[7] Discutiremos um exemplo concreto na seção seguinte, na esperança de tornar o problema mais claro.

**O método meta-popperiano**

Alguém poderia dizer "viu só? É por *isso* que não convidamos o pessoal da metafísica para entrar no laboratório!". Nesse sentido, a metafísica traria mais questões do que respostas. Essa afirmação não é, digamos, *totalmente* equivocada. Afinal, se não há esperanças de ancorar a metafísica na ciência, no sentido de extrairmos a metafísica *da* ciência, a subdeterminação metafísica é algo que podemos tomar por garantido. Todavia, a subdeterminação é um daqueles fatos da vida com os quais devemos aprender a conviver, seja na ciência, seja na ontologia, seja na metafísica. A questão central passa a ser como fazer isso de forma a minimizar os danos, e como vislumbrar quais alternativas possuem mais chances de serem bem sucedidas.

De fato, há um fio de esperança que promete arrumar uma parte dessa situação embaraçosa. Existe um método para descartarmos quais as alternativas *erradas*, ou melhor dizendo: os perfis metafísicos *incompatíveis* com certas teorias. Isto é, nem *tudo* vale, e alguns perfis metafísicos não estão à altura da tarefa. Por mais que a metafísica flutue livremente da ciência, enquanto disciplina, não é *qualquer* metafísica que pode interpretar *qualquer* entidade de *qualquer* teoria científica. Podemos não saber qual é a *correta*, pois isso envolve diversas questões relacionadas à natureza da subdeterminação. Esse método foi originalmente chamado de meta-popperiano (Arenhart, 2012), mas recentemente também tem sido chamado de "estórias metafísicas indisponíveis" (Arroyo, 2020a; 2020b).

Antes de apresentar o método, contudo, é oportuno prepararmos um terreno para a sua aplicação nas interpretações da mecânica quântica.[8] Um dos casos mais fáceis da aplicação desse método é a limitação do rol de possibilidades metafísicas da interpretação que chamaremos de "interpretação da consciência causal" (Arroyo; Arenhart, 2019), que talvez seja uma das interpretações da mecânica quântica dentre as mais rejeitadas por parte de físicos e filósofos (Schlosshauer, Kofler, e Zeilinger, 2013), mas, ainda assim, uma opção disponível para interpretar os fenômenos quânticos (de Barros e Oas, 2017; de Barros e Montemayor, 2019), com proponentes ativos até hoje (Stapp, 2009).

---

[8] Não discutiremos aqui se o termo "interpretação" é adequado ou não. Ainda que tenhamos optado por manter a terminologia usual, apontamos que há uma série de problemas, recentemente explicitados, principalmente, por Maudlin (2019), Dürr e Lazarovici (2020), e Arroyo e da Silva (2021).

Muito brevemente, e em termos não técnicos, pode-se dizer que essa interpretação da consciência causal recebe seu nome por meio de um processo ad-hoc que introduz na solução do problema da medição na mecânica quântica. O raciocínio é o seguinte: os sistemas físicos obedecem às leis da mecânica quântica; um sistema físico pode estar em uma superposição dos autoestados do observável sendo medido, mas sempre observamos o autovalor correspondendo a um único autoestado (nunca autoestados superpostos), associado a um estado de ponteiro do aparelho de medição; mas dado que o aparelho é também um sistema físico, ele deveria obedecer às leis da mecânica quântica; por conseguinte, deve também ser descrito por uma superposição de estados de ponteiro. Então o problema é colocado da seguinte maneira: como podemos observar somente estados definidos, e não em superposição? A solução da consciência causal consiste em dizer que sempre que os sistemas físicos interagem com a consciência de um observador,[9] eles *colapsam* para um de seus estados possíveis, de modo que a consciência é o agente causal desse colapso. Para visualizarmos, imagine uma situação em que o aparelho medidor pudesse ler os estados 'para cima' e 'para baixo', com igual probabilidade. A descrição quântica, antes dessa interação com sistemas conscientes (não sujeitos às leis da mecânica quântica) seria a soma dos dois estados, e a interação com a consciência causal resultaria em apenas um estado.

Como realistas rasos e profundos reagiriam a isso? Realistas do tipo 'raso' naturalmente focam em soluções para a subdeterminação: ora, se podemos extrair que a consciência existe, de acordo com essa interpretação, mas também podemos extrair outros comprometimentos ontológicos diferentes de outras interpretações, o problema é a subdeterminação pelos dados que a própria ciência enfrenta. Nesse caso, o realista raso iria se preocupar com argumentos que pudessem eventualmente favorecer uma interpretação da mecânica quântica para que possa clamar uma atitude realista ao seu catálogo ontológico correspondente com maior justificação epistêmica. Por exemplo, no caso de alguém ser realista (raso) quanto à interpretação da consciência causal referida acima, a atitude mais natural é buscar favorecer essa interpretação por elementos teóricos na física (e.g., buscando justificações para a noção de 'colapso') ou na filosofia (parcimônia ontológica em relação a outras interpretações, como a dos muitos mundos). No entanto, como o debate de quase um século tem mostrado, essa não é uma tarefa fácil — e nem mesmo está próxima de ser realizada com sucesso (Dürr e Lazarovici, 2020). Realistas do tipo 'profundo', no entanto, não se contentam com isso. Motivados pela ideia segundo a qual, sem uma resposta para o Desafio de Chakravartty (viz. sem um perfil metafísico) o realismo científico não é legítimo, resolver a subdeterminação teórica é apenas um primeiro passo na busca para o realismo. Por isso, buscam alternativas para entender, metafisicamente, o que são essas entidades com as quais as teorias se comprometem, pois somente assim poderiam adotar uma postura legitimamente realista em relação a essas mesmas teorias. Então, para ficar no exemplo da consciência causal, buscam teorias metafísicas que possam 'vestir' essa consciência: dualismo, emergentismo, fisicalismo, etc. Mas, como já mencionamos, essas alternativas metafísicas parecem se proliferar indefinidamente. Em outras palavras: para o realista de tipo raso, a subdeterminação é do tipo teórico, ocorrendo entre diferentes interpretações possíveis e suas diferentes mobílias para o mudo. O realista profundo, no entanto, *acrescenta* a esse problema mais um: uma vez que se aceite uma interpretação e sua correspondente ontologia, como a interpretação que atribui um poder causal para a consciência, teremos uma nova camada de subdeterminação, desta vez acerca da atribuição de um perfil metafísico para essa consciência com poderes causais (já que, lembre, ele exige que se vá mais a fundo). O que podemos fazer?

---

[9] O termo 'consciência' é talvez um dos termos mais mal-definidos na mecânica quântica (Bueno, 2019), e seus maus usos foram explorados, dentre outras coisas, por um fenômeno cultural conhecido como "misticismo quântico" (Pessoa Junior, 2011).

Com essa situação em mente, passemos à apresentação do método. Podemos até pensar no anúncio: *Trazemos sua metafísica incompatível em três passos!* No primeiro passo, que podemos chamar de *quineano*, identificamos quais os comprometimentos ontológicos de uma teoria científica em questão, isto é, "[…] o que, de acordo com essa teoria, existe" (Quine, 1951, p. 65). Por exemplo, em algumas interpretações da mecânica quântica, a consciência humana é usada para resolver um problema nos fundamentos da teoria, e para isso funcionar, essa consciência deve ter poderes causais. Então encontramos no nosso caderno de ontologia essa entidade, a "consciência", e marcamos o quadradinho dela como "existente".

No segundo passo, que podemos chamar de *Viking*, pegamos nosso barco e vamos atrás dos vilarejos da metafísica, e pegamos todos os conteúdos produzidos sobre essa entidade. No nosso exemplo, essa entidade é a "consciência". Então vamos de Descartes a Dennett, pilhando e saqueando a literatura numa analogia terrível,[10] mas que ao menos é uma forma de ilustrar o fato de que *não estamos produzindo conteúdo metafísico algum!*

No terceiro passo, que poderíamos chamar de *popperiano*, nós avaliamos os espólios obtidos no passo anterior. Isto é, nós checamos a incompatibilidade das teorias metafísicas obtidas com as entidades ontológicas obtidas, e vemos se há alguma restrição ontológica ou científica para vestirmos essas entidades metafisicamente de certo modo. É aqui que a parte negativa acontece, acenando para uma atitude *moderadamente naturalizada*. Pode ser que nem toda metafísica seja compatível com as entidades obtidas, e, nesse caso, a ciência acaba sendo privilegiada. Isto é, em caso de incompatibilidade entre uma teoria metafísica extraída de livros de filosofia, e uma ontologia extraída da própria ciência, abandonamos a teoria metafísica.

Note que a ordem dos passos é crucial, pois nós precisamos das entidades para termos *ao que* dar um perfil metafísico. Uma das formas é obter as entidades pelo comprometimento ontológico. Então o "passo Viking" estaria condicionado às entidades obtidas em dada teoria em questão, e é aí que podemos ver se de fato são incompatíveis ou não (e é aqui que temos uma conexão com o Desafio de Chakravartty, claro). De outro modo ficaria difícil discutirmos sobre a natureza metafísica de entidades as quais nem mesmo admitimos na ontologia.

---

[10] Posteriormente, French (2020) reconhece, influenciado por McKenzie (French e McKenzie, 2012) que essa analogia era, de fato, muito cruel. Esse é o motivo pelo qual essa abordagem hoje é identificada como abordagem da *Caixa de ferramentas* ("Toolbox").

Isso pode ficar mais claro se voltarmos ao nosso exemplo da consciência causal na mecânica quântica: por que discutir sobre a natureza da consciência, caso nem mesmo admitamos essa entidade na nossa ontologia? Por isso o primeiro passo é essencial: no passo quineano, vimos que, em ao menos uma interpretação da mecânica quântica, a consciência precisa existir, e precisa existir com *reais poderes causais*. A partir do momento em que se adota ou assume essa versão da teoria, mesmo que provisoriamente, uma metafísica dualista parece ser bem acomodada aqui, a princípio podendo interpretar essa entidade, a consciência, como *res cogitans*, por exemplo.[11] O mesmo não acontece com uma metafísica fisicalista eliminativista, na qual a consciência é desprovida de qualquer tipo de realidade. É importante notar que não se trata de uma questão de dizer que o fisicalismo é uma metafísica errada, e que o dualismo é uma metafísica correta: é uma simples questão de *incompatibilidade* metafísica com os ditames teóricos e ontológicos de uma teoria em questão. No caso, o eliminativismo é uma fonte de perfil metafísico *incompatível* com essa interpretação da mecânica quântica. Assim, deve ficar claro: o método não é capaz de falsear ou selecionar uma interpretação da mecânica quântica (ele não opera, portanto, ao nível do realismo raso); o que o método faz é mostrar que não podemos ter as duas coisas juntas: o eliminativismo (e também o cartesianismo, ver nota 8) *e* a interpretação da consciência causal da mecânica quântica. O mesmo caso acontece quando tomamos o caso da interpretação fisicalista eliminativista que vigora como um padrão acerca da natureza da consciência na neurociência cognitiva em conjunto com essa particular interpretação da mecânica quântica da consciência causal: não podemos ter as duas ao mesmo tempo devido às suas incompatibilidades metafísicas (e.g. como pode uma consciência causar o colapso se ela nem mesmo existe?). Note que apesar de se tratar de uma incompatibilidade explícita entre uma concepção de consciência proposta por uma teoria científica, a neurociência cognitiva, nossa ênfase recai no fato de que se tem uma visão da natureza da consciência (um perfil metafísico) que é desautorizada pela interpretação da mecânica quântica particular que estamos tratando. Vale a pena mencionar que é precisamente nesse ponto que uma *veia naturalista* pode ser encontrada no referido método meta-popperiano: no caso de termos um conflito entre uma teoria científica, por um lado, e uma teoria metafísica, por outro, quem sai é a teoria metafísica! Isto é, diante do conflito entre metafísica e ciência, parecem existir razões pragmáticas e epistêmicas para modificarmos a metafísica ao invés de modificarmos a física. Isso não é de maneira alguma consensual, haja vista que diversos filósofos poderiam objetar a esse ponto, pedindo justamente o contrário: no caso de um conflito entre a ciência e a filosofia, pior para a ciência (Lowe, 1998; Benovsky, 2016). De outro ponto de vista, os naturalistas com os quais estivemos dialogando até agora tenderiam para que as teorias filosóficas sejam abandonadas (quiçá com a disciplina toda *descontinuada*, tal como sugerem Ladyman e Ross, 2017). O meta-popperiano situa-se no exato meio entre esses dois extremos quando afirma, no fundo, que não podemos ter as duas coisas: uma teoria metafísica e uma teoria científica *quando o método acusa sua incompatibilidade*; nesse caso, como já dissemos, a teoria metafísica é quem tem que ir. O privilégio epistêmico, nesse caso, é da ciência. Novamente, é importante enfatizar que a ciência possui um privilégio epistêmico do ponto de vista do meta-popperiano justamente pelo fato de que estamos buscando uma relação entre metafísica e ciência, através da qual a metafísica possa se beneficiar da ciência. Certamente, para aqueles que negam esse privilégio epistêmico da ciência, ou que duvidam que a metafísica precise de algum tipo de respaldo científico para garantir sua legitimidade, essa relação sequer precisa ser buscada. Mas esse é outro tipo de debate, com o qual não estamos preocupados aqui.

---

[11] Ainda assim, vale dizer que nem todo tipo de dualismo é compatível com essa interpretação da mecânica quântica. Seguindo a taxonomia oferecida por Rodrigues (2014, p. 203), o dualismo cartesiano seria classificado como um "dualismo puro não espacial interacionista teísta forte". No entanto, como mostraram Arroyo e Arenhart (2019, pp. 37–38), as únicas formas de dualismo compatíveis com a interpretação da consciência causal da mecânica quântica são as versões fortes dos dualismos naturalistas e interacionistas — de modo que outros perfis metafísicos dualistas para a consciência são descartados pelo método meta-popperiano como incompatíveis com a referida interpretação da mecânica quântica.

*Essa* busca pela incompatibilidade, com ênfase na inadequação de teorias metafísicas obtida a partir de teorias científicas seria, na nossa opinião, uma forma de se explicitar uma relação produtiva entre ciência e metafísica, na medida em que mostra de fato como um relacionamento produtivo entre as duas áreas se dá na prática, e como essa interação se mostra mutuamente benéfica: do lado da metafísica, podemos ver quais as teorias metafísicas que são incompatíveis com certas teorias científicas; do lado da ciência, podemos não saber qual a forma correta de interpretarmos as entidades com as quais se compromete, mas podemos saber quais as formas *incorretas* — o que no estado atual do debate, já é uma grande coisa.

Esse método tem sido aplicado aos problemas metafísicos em relação à individualidade das partículas quânticas (Arenhart, 2012; 2017), a outras interpretações da mecânica quântica como a interpretação dos muitos mundos (Arroyo, 2020a), e a interpretação fenomenológica (Arroyo; Nunes Filho, 2018), e, recentemente, à epidemiologia e o novo Coronavírus (Arroyo, 2020b). No entanto, ainda há muito o que se fazer.

**Conclusão**

A mensagem para levar pra casa é a seguinte. Se admitirmos que a metafísica é importante para a descrição científica (e isso é um grande 'se'!), somos deixados com uma miríade de metafísicas possíveis para interpretar as entidades com as quais as teorias científicas se comprometem para que funcionem corretamente. A alternativa que propusemos através do método meta-popperiano é a de avaliar quais perfis metafísicos não são de fato alternativas metafísicas, observando as restrições teóricas e ontológicas fornecidas pelas próprias teorias científicas. No caso da mecânica quântica, note, há ainda um 'se' anterior: a metafísica, ao depender de uma escolha de ontologia, também depende da escolha de uma interpretação, e *esse* problema também fica subdeterminado pela mecânica quântica. Todavia, o problema da subdeterminação da escolha de uma interpretação parece relacionado com a escolha de uma versão apropriada da mecânica quântica, um problema para os físicos, enquanto que a escolha de uma metafísica apropriada, a ser colocada sobre uma interpretação, é um problema para os filósofos da ciência que desejam ser realistas profundos. Todavia, uma vez que se escolhe uma interpretação, ela delimita o campo de possibilidades abertas para as metafísicas que podem ser empregadas junto a ela.

Esse é um passo em direção a uma descrição metafísica que, em certa medida, é informada pela ciência, como naturalistas gostariam. Por outro lado, agrada também os metafísicos da ciência, já que lida com uma certa autonomia da metafísica enquanto disciplina: se a metafísica não é *extraída* da ciência, o espaço para a criatividade e para a exploração metafísica está garantido! Mas nem tudo vale: podemos não saber quais as alternativas corretas, mas ao menos temos elementos suficientes em mãos, com o supracitado método, para sabermos quais não são, objetivamente, as alternativas metafísicas a serem perseguidas para interpretar a ciência atual.

Por fim, defendemos que a metafísica não é assim tão livre como gostariam os Vikings. Ela 'flutua' em relação à ciência, mas como uma pipa: há uma linha que a segura, que é a ciência. Assim, sempre poderão haver muitas pipas no ar, e o método meta-popperiano nos indica quais as "pipas sem corda". Isso parece ser, no momento, o melhor que podemos fazer.

**Referências**